\documentclass[11pt,a4paper]{article}

\usepackage{jcappub}
\usepackage{epstopdf}
\DeclareGraphicsExtensions{.pdf,.png,.jpg}
\usepackage{amsmath}
\usepackage{amsfonts}
\usepackage{amssymb}
\usepackage{url}

\title{Cosmology with matter diffusion}

\author[a]{Simone~Calogero}
\author[b]{and Hermano~Velten}

\affiliation[a] {Mathematical Sciences, Chalmers University of Technology, Gothenburg University,\\
S-412~96 Gothenburg,
Sweden}
\affiliation[b]{Universidade Federal do Esp\'irito Santo, Av. Fernando Ferrari, Goiabeiras, Vit\'oria, Brasil}
\emailAdd{calogero@chalmers.se}
\emailAdd{velten@cce.ufes.br}

\abstract{We construct a viable cosmological model based on velocity diffusion of matter particles. In order to ensure the conservation of the total energy-momentum tensor in the presence of diffusion, we include a cosmological scalar field $\phi$ which we identify with the dark energy component of the universe. The model is characterized by only one new degree of freedom, the diffusion parameter $\sigma$. The standard $\Lambda$CDM model can be recovered by setting $\sigma=0$. If diffusion takes place ($\sigma >0$) the dynamics of the matter and of the dark energy fields are coupled. We argue that the existence of a diffusion mechanism in the universe may serve as a theoretical motivation for interacting models. We constrain the background dynamics of the diffusion model with Supernovae, H(z) and BAO data. We also perform a perturbative analysis of this model in order to understand structure formation in the universe. We calculate the impact of diffusion both on the CMB spectrum, with particular attention to the integrated Sachs-Wolfe signal, and on the matter power spectrum $P(k)$. The latter analysis places strong constraints on the magnitude of the diffusion mechanism but does not rule out the model.}


\begin{document}
\maketitle


\section{Introduction}

The recent release of the PLANCK data \cite{planck16}, the european space agency's satellite designed to produce the most accurate observation of the Cosmic Microwave Background (CMB) sky ever, has confirmed that the $\Lambda$CDM model is our best, and at the same time the simplest, general relativity based cosmological model. The term $\Lambda$ refers to a dark energy component in the form of a cosmological constant whilst CDM stands for Cold Dark Matter. Together they form the dark sector which corresponds to $\sim 95\%$ of today's cosmic energy budget with the remaining fraction left to the baryonic matter $\sim 5\%$ and an almost negligible radiation background. Actually, the acronym $\Lambda$CDM represents only a simplified view of the standard cosmological model which relies on many other pillars like, for example, the large scale homogeneity and isotropy of the universe, the existence of an inflationary epoch which seeds structure formation, the perfect fluid behavior of the cosmic components and the adiabaticity of the cosmic medium. The latter is intrinsically related to the fact that different cosmic fluids are assumed to obey separately the energy-momentum conservation. The adiabaticity is also guaranteed because it is assumed that no dissipative effects take place in the universe.

Despite of the remarkable agreement of the $\Lambda$CDM model with the observations, not only of CMB, but also of Supernovae \cite{suzuki}, BAO \cite{sdss, wiggle, 6dfgrs}, and other data sets (see \cite{reviewDE} for a recent review), alternative scenarios are still in the game. In general terms, the other approaches are based on either the use of modified gravity theories \cite{modgrav} or the use of non-standard cosmic components, e.g, models where the dark energy is a quintessence \cite{quintessence} or a phantom  field~\cite{pha} or when the dark energy's equation of state is a time evolving quantity \cite{cpl}. Another interesting scenario is provided by the class of interacting models \cite{int1,int2,int3,int4,int5}, i.e., models that assume a phenomenological coupling between the dark components in the form of energy flowing from dark energy to dark matter or vice versa (the direction of the energy flux depends only on the sign of the interacting term). The energy-momentum conservation of the individual cosmic components does not hold anymore, but the interaction is chosen in such a way that the total energy-momentum tensor is divergence-free, which is necessary in view of the Einstein equations and the Bianchi identities. The main physical motivation for studying interacting models is that the coupling between dark energy and dark matter could resolve the coincidence problem \cite{coincidence}, namely the puzzling circumstance that dark energy and dark matter have similar energy densities today while, according to the $\Lambda$CDM model, they differed of several orders of magnitude in the distant past. 

As very little is known on the nature of the dark components, the interacting models have been introduced so far only on a pure phenomenological, rather than physical, basis. In this work we propose a possible explanation for the interaction between dark matter and dark energy, which is grounded on a physical mechanism that is ubiquitous in nature: diffusion.  Our starting point here is the assumption that the matter particles (baryonic and dark) undergo velocity diffusion in an expanding, homogeneous and isotropic universe. It turns out that an interaction between dark energy and dark matter results as a natural consequence of the relativistic diffusion mechanism. We remark that dissipative effects are not very unlikely to occur in the universe \cite{schwarz2012}. Hence, the introduction of dissipative mechanisms (in our case, the matter particles velocity diffusion) represents a step towards a more realistic description of the dynamics of the universe than the one provided by perfect fluids.

We start by developing a cosmological model which admits matter diffusion \cite{Ca1}, carrying over the findings of Ref. \cite{Ca2}. In the latter reference it is shown that matter diffusion can only take place if spacetime is permeated by a field that balances the energy gained by the matter particles due to diffusion. The simplest and more reasonable choice for this new field is that of a cosmological scalar field (the $\phi$ field) which we identify here with the dark energy component of the universe. Our model is composed by a matter like component and the $\phi$ field, giving rise to the $\phi$CDM model. In comparison to the $\Lambda$CDM model, the $\phi$CDM model has only one extra degree of freedom, the diffusion parameter $\sigma >0 $. The density associated to the $\phi$ field is a time evolving quantity if diffusion takes places and $\Lambda$CDM can be achieved by setting $\sigma=0$.

In order to construct a viable model we will constrain the background dynamics with recent observational data sets like Supernovae, H(z) data and Baryonic Acoustic Oscillations (BAO) data. We also make use of the cosmological perturbation theory at linear order to understand how the presence of diffusion affects structure formation. We show that the Cosmic Microwave Background (CMB) data and the matter power spectrum are very sensitive probes for the diffusion mechanism because they place the strongest constraints on the diffusion parameter.

The precise outline of the paper is as follows. In the next section we briefly recall the main ingredients of diffusion theory in the context of general relativity. The diffusion dynamics in an expanding, homogeneous and isotropic background are studied in Sec.~\ref{2.1}, the particular case of a spatially flat universe being the $\phi$CDM model introduced in Sec.~\ref{2.2}. We develop a (linear, scalar) cosmological perturbation theory for the diffusion model in Sec.~\ref{cosmopert} and the solutions for the perturbed quantities are studied in Sec.~\ref{sollin}. Observational constraints on the diffusion parameter from background data and large scale structure (LSS) formation analysis are studied in Sec.~\ref{CompObsBack} and Sec.~\ref{CompObsPert}, respectively.

\section{Cosmological models with diffusion}\label{cosmomodel}
We begin this section by discussing the general theory for the diffusion dynamics of matter in general relativity, and then we specialize to the models applied in Cosmology. 

Let $T_{\mu\nu}$ and $J^\mu$ be the energy-momentum tensor and the current density of some matter distribution in a spacetime $(M,g)$. The matter is said to undergo {\it microscopic} (or molecular) velocity diffusion in a cosmological scalar field $\phi$ if the following equations hold:
\begin{equation}\label{EinsteinEq}
R_{\mu\nu}-\frac{1}{2}g_{\mu\nu}R+\phi g_{\mu\nu}=T_{\mu\nu},
\end{equation} 
\begin{equation}\label{diffusion}
\nabla_\mu T^{\mu\nu}=\sigma J^\mu,
\end{equation}
\begin{equation}\label{consJ}
\nabla_\mu J^\mu=0.
\end{equation}
As usual, $R_{\mu\nu}$ denotes the Ricci curvature of the metric $g$ and $R=g^{\mu\nu}R_{\mu\nu}$. Eq.~\eqref{diffusion} is the {\it macroscopic} diffusion equation.  The constant $\sigma>0$ is the diffusion constant. The value $3\sigma$ measures the energy transferred from the scalar field to the matter per unit of time due to diffusion. We use units $8\pi G=c=1$. 

Our interpretation of the system~\eqref{diffusion} stems from the kinetic model introduced in~\cite{Ca1}. We emphasize that, according to the kinetic formulation given in~\cite{Ca1}, it is the microscopic velocity of the particles that is subject to diffusion and not the macroscopic four-velocity field of matter. Hence, in particular, diffusion takes place on the tangent bundle of spacetime (the phase space) and there is no danger of Lorentz invariance breaking in the model.

Taking a covariant divergence of~\eqref{EinsteinEq}, using~\eqref{diffusion} and the Bianchi identity $\nabla^\mu (R_{\mu\nu}-\frac{1}{2}g_{\mu\nu}R)=0$, we obtain the following evolution equation for $\phi$:
\begin{equation}\label{phieq}
\nabla_\mu\phi=\sigma J_\mu.
\end{equation}

If the matter undergoing diffusion is a perfect fluid, the energy-momentum tensor and the current density are given by
\begin{equation}\label{TandJfluid}
T_{\mu\nu}=\rho u_\mu u_\nu +p(g_{\mu\nu}+u_\mu u_\nu),\quad J^\mu=n u^\mu, 
\end{equation}
where $\rho$ is the rest-frame energy density, $p$ the pressure, $u^\mu$ the 4-velocity and $n$ the particle number density of the fluid. For a perfect fluid, Eq.~\eqref{phieq} reads
\begin{equation}\label{phieqfluid}
\nabla_\mu\phi=\sigma n u_\mu,
\end{equation}
while Eq.~\eqref{consJ} becomes 
\begin{subequations}\label{fluideqs}
\begin{equation}\label{Jeqfluid}
\nabla_\mu (n u^\mu)=0.
\end{equation}
Moreover projecting~\eqref{diffusion} into the direction of $u^\mu$ and onto the hypersurface orthogonal to $u^\mu$ we obtain
\begin{equation}\label{conteq}
\nabla_\mu(\rho u^\mu)+p\nabla_\mu u^\mu=\sigma n,
\end{equation}
\begin{equation}\label{euler}
(\rho+p)u^\mu \nabla_\mu u^\nu+u^\nu u^\mu\nabla_\mu p + g^{\mu\nu}\nabla_\mu p=0.
\end{equation}
\end{subequations}   
In the absence of diffusion, i.e., when $\sigma=0$, the cosmological scalar field is constant throughout spacetime and~\eqref{EinsteinEq} reduces to the Einstein equation with cosmological constant, while~\eqref{fluideqs} reduce to the relativistic Euler equations for perfect fluids. 

Finally we remark that it is possible to consider different models, other than a cosmological scalar field, for the background medium in which diffusion takes place (e.g., a perfect fluid, see~\cite[Sec.~4]{Ca2}). It is currently unknown whether any such model admits a Lagrangian formulation.

\subsection{Cosmological dynamics with diffusion}\label{2.1}

The cosmological principle, asserting that the universe is homogeneous and isotropic, assigns a special role to the cosmological models based on the class of FLRW metrics. CMB observations show a high degree of homogeneity and isotropy at large scales, but they also reveal the existence  of fluctuations of order $10^{-5}$ in the CMB temperature already at $z\sim 1100$. Such tiny fluctuations evolved in time via gravitational instability and formed the structures visible in the universe. Therefore a complete description of the universe dynamics must include the large scale background expansion as well as the evolution of the matter fluctuations that give rise to structures. This is the task of cosmological perturbations theory, which we apply to the diffusion model in Section~\ref{cosmopert}. 
Prior to this, we undertake the important analysis of the diffusion model under the assumption of spatial homogeneity and isotropy, i.e., we consider a spacetime with the Robertson-Walker metric
\begin{equation}\label{RW}
ds^2=-dt^2+a(t)^2\left[\frac{dr^2}{1-k r^2}+r^2d\Omega^2\right],\quad k=0,\pm 1.
\end{equation}
In the spatially flat case $k=0$ we may introduce a cartesian system of coordinates such that 
\begin{equation}\label{spatiallyflatRW}
k=0:\quad ds^2=-dt^2+a(t)^2(dx^2+dy^2+dz^2),\quad\text{and}\ a_0:=a(0)=1.
\end{equation}
The symmetry assumption forces the fluid to be comoving, i.e., $u^\mu=(1,0,0,0)$. Moreover the remaining fluid variables depend only on $t$. Hence~\eqref{Jeqfluid} entails
\begin{equation}\label{nfluid}
n(t) a(t)^3=const.\ \Rightarrow\ n(t)=\frac{n_0}{a(t)^3}.
\end{equation}
A subscript 0 indicates that the corresponding quantity is evaluated at the present value of the cosmological time, which we take at $t=0$.
Eq.~\eqref{conteq} becomes
\begin{equation}\label{conteqhom}
\dot{\rho}+3H(\rho+p)=\sigma n_0a^{-3},
\end{equation}
where 
\begin{equation}\label{H}
H=\frac{\dot{a}}{a}
\end{equation}
is the Hubble function. Eq.~\eqref{euler} is identically satisfied. The cosmological scalar field equation~\eqref{phieqfluid} reduces to
\begin{equation}\label{phidot}
\dot{\phi}=-\sigma n_0 a^{-3}.
\end{equation} 
The only non-trivial, independent Einstein equation is the Hamiltonian constraint
\begin{equation}
H^2=\frac{1}{3}(\rho+\phi)-\frac{k}{a^2}.\label{constraint}
\end{equation}

The initial data set consists of $(a_0,H_0,\rho_0,\phi_0)$  of {\it positive} numbers such that~\eqref{constraint} is satisfied at time $t=0$, i.e.,
\[
H_0^2=\frac{1}{3}(\rho_0+\phi_0)-\frac{k}{a_0^2}.
\]

In the rest of this section we assume a linear equation of state between the pressure and the energy density: 
\begin{equation}\label{eqofstate}
p=(\gamma-1)\rho,\quad 2/3<\gamma<2.
\end{equation}
The bounds on $\gamma$ ensure that the cosmological fluid satisfies the dominant and the strong energy condition.
%
%
As shown in~\cite{Ca2}, the asymptotic behavior for large times in the past and future directions of solutions to the system~\eqref{conteqhom}-\eqref{eqofstate} is rather sensitive to the choice of initial data and in fact each of the following type of solutions have been found numerically:
\begin{itemize}
\item[(a)] Solutions with singularity in the past (Big-Bang) and in the future (Big-Crunch)
\item[(b)] Solutions with singularity in the past, which are singularity free in the future
\item[(c)] Solutions with no singularity in the past  
\end{itemize}
Solutions of type (c), also called {\it bouncing models}, are excluded by observations~\cite{reviewDE} and will not be considered further in this paper.  For the remaining solutions, $a(t)$ is decreasing toward the past and the time variable $t$ can be replaced by the cosmological redshift variable
\begin{equation}
z(t)=\frac{a_0}{a(t)}-1,\quad z>0 \quad \text{(i.e., $t<0$).}
\end{equation}

Introducing the dimensionless variables
\begin{eqnarray}
\Omega_{m}(z)=\frac{\rho(z)}{3H^2_0}, \hspace{1cm}\Omega_{\phi}(z)=\frac{\phi(z)}{3H^2_0},\hspace{1cm} E(z)=\frac{H(z)}{H_0} \nonumber
\end{eqnarray}
and the dimensionless parameters
\begin{equation}
\tilde{\sigma}=\frac{\sigma n_0}{3a_0^3H^3_0}, \quad K=\frac{k}{H_0^2a_0^2}
\end{equation}
we can rewrite~\eqref{conteqhom},~\eqref{phidot} and~\eqref{constraint} as
\begin{equation}\label{dimensionlesseq1}
\frac{d\Omega_m(z)}{dz}=\frac{3\gamma\Omega_m(z)}{1+z}-\tilde{\sigma}\frac{(1+z)^2}{E(z)},
\end{equation}
\begin{equation}\label{dimensionlesseq2}
\frac{d\Omega_\phi(z)}{dz}=\tilde{\sigma}\frac{(1+z)^2}{E(z)},
\end{equation}
\begin{equation}\label{dimensionlesseq3}
E(z)=\sqrt{\Omega_m(z)+\Omega_\phi(z)-K(1+z)^2}.
\end{equation} 
The previous system describes the past (i.e., $z>0$) of diffusion cosmological models possessing a Big-Bang singularity. If spacetime is forever expanding in the future, then the system~\eqref{dimensionlesseq1}-\eqref{dimensionlesseq3} gives also a complete description of the cosmological model in the future direction. An example of solution defined for all $z\in (-1,+\infty)$ is given by
\begin{subequations}\label{solution}
\begin{equation}
\Omega_m(z)=\frac{2\beta_k}{3\gamma-2}(1+z)^2,\quad \Omega_{\phi}(z)=\beta_k(1+z)^2,
\end{equation}
where $\beta_k$ is the real root of the equation
\begin{equation}\label{polynomial}
\frac{12\gamma}{3\gamma-2}\beta^3-4K\beta^2-\tilde{\sigma}^2=0.
\end{equation}
\end{subequations}
Note that in the limiting case $\tilde{\sigma}=0$ the roots of~\eqref{polynomial} are 
\[
\beta=\frac{(3\gamma-2)}{3\gamma},\ \text{for }K=1;\quad \beta=0,\ \text{for }K=-1,0,1.
\]
The roots for $K=0$ and $K=-1$ correspond to vacuum solutions of the diffusion-free Einstein equations, namely the Minkowski spacetime for $K=0$ and the Milne universe for $K=-1$, while there is no solution associated to the the roots for $K=1$. Hence the solution~\eqref{solution} of~\eqref{dimensionlesseq1}-\eqref{dimensionlesseq2} is a pure diffusion one.

\subsection{The $\phi$CDM model} \label{2.2}
We call $\phi$CDM model the cosmological model which is obtained by setting $p=k=0$ in the system~\eqref{conteqhom}-\eqref{constraint}. This model reduces exactly to the standard $\Lambda$CDM model for $\sigma=0$, i.e., in the absence of diffusion. We remark that the equation of state $p=0$, i.e., $\gamma=1$, corresponds to a dust fluid, which is the expected behavior of dark matter and baryons. According to the standard cosmological model, the pressureless matter component is responsible for $\sim 30 \%$ of today's cosmic energy budget. The remaining $\sim 70\%$ would be in the form of a dark energy, which is described by the cosmological constant $\Lambda$ in the $\Lambda$CDM model, and by the cosmological scalar field $\phi$ in the $\phi$CDM model. 

Since $a_0=1$ and $K=0$ for the $\phi$CDM model, the past of the solutions possessing a Big-Bang singularity is described by the system 
\begin{equation}\label{dimensionlesseq1new}
\frac{d\Omega_m(z)}{dz}=\frac{3\Omega_m(z)}{1+z}-\tilde{\sigma}\frac{(1+z)^2}{E(z)},
\end{equation}
\begin{equation}\label{dimensionlesseq2new}
\frac{d\Omega_\phi(z)}{dz}=\tilde{\sigma}\frac{(1+z)^2}{E(z)},
\end{equation}
\begin{equation}\label{dimensionlesseq3new}
E(z)=\sqrt{\Omega_m(z)+\Omega_\phi(z)},
\end{equation} 
where 
\begin{equation}
z=a^{-1}-1,\quad \tilde{\sigma}=\frac{\sigma n_0}{3H^3_0},
\end{equation}
For $\tilde{\sigma}=0$ the solution is given by the $\Lambda$CDM model:
\[
\Omega^{(0)}_m(z)=\Omega^{(0)}_{m0}(1+z)^3,\quad\Omega_\phi^{(0)}(z)=\Omega_{\phi0}^{(0)}=1-\Omega^{(0)}_{m0}.
\]
%

\begin{figure}
\begin{center}
\includegraphics[width=0.435\textwidth]{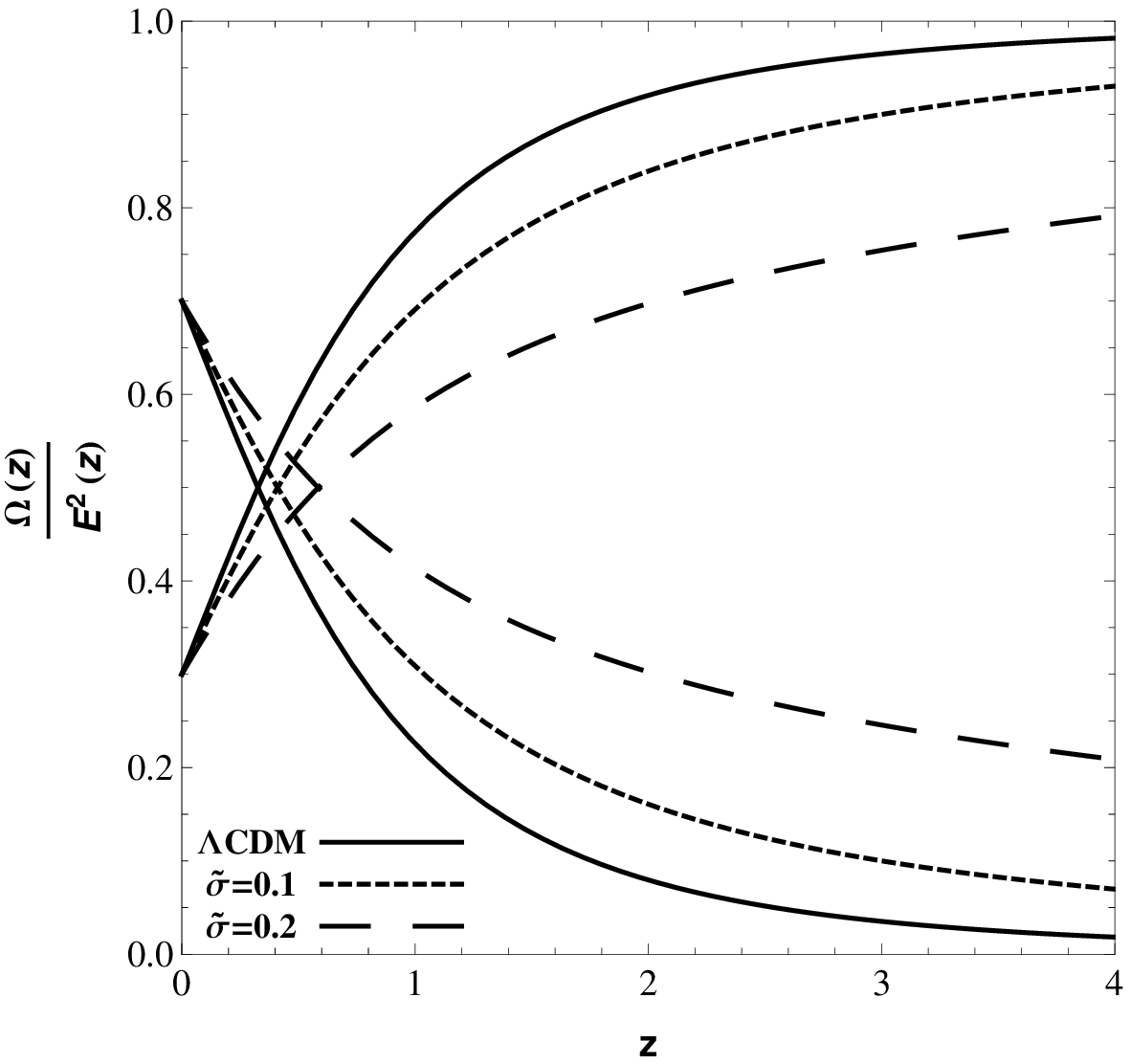}
\includegraphics[width=0.43\textwidth]{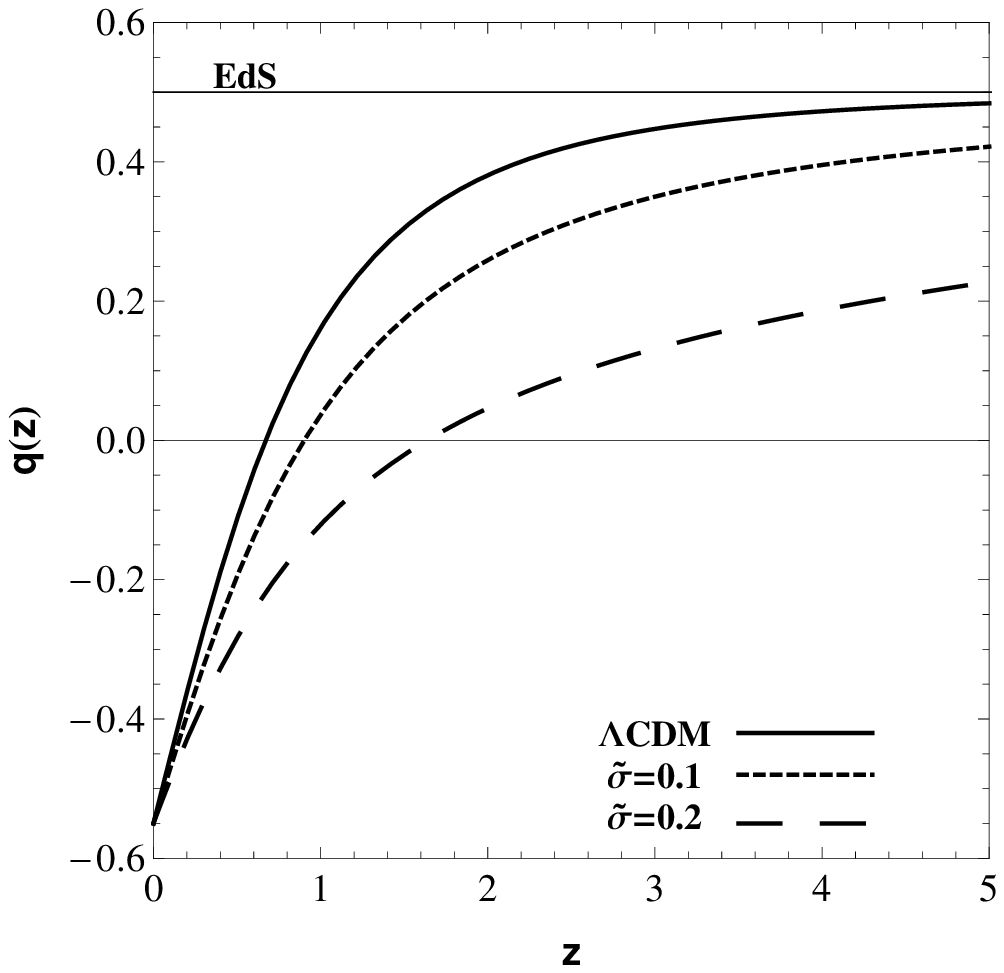}	
\caption{{\it Left Panel:} Evolution of the fractional density parameters. {\it Right panel:} Evolution of the deceleration parameter (right). Both assume $\Omega_{m0}=0.3$. In the right panel, the solid horizontal line at the value $q=0.5$ describes the Einstein-de-Sitter (EdS) model and solid horizontal line at the value $q=0$ denotes the transition to the accelerated expansion. }
\label{fig1}
\end{center}
\end{figure}

In the left panel of Fig.~\ref{fig1} we see the behavior of the parameters $\Omega_{m}(z)/E^2(z)$ (increasing curves) and  $\Omega_{\phi}(z)/E^2(z)$ (decreasing curves) for the matter and the dark energy fluids. The solid lines represent the standard $\Lambda$CDM model, while dashed lines depict the behavior of the $\phi$CDM model for $\sigma=0.1$ and $\sigma=0.2$. The right panel corresponds to the deceleration parameter $q(z)=-1-\dot{H}/H^2$ for the same model parameters. From the latter picture we see that the phase of accelerated expansion of the universe ($q<0$) begins earlier  in the presence of diffusion. 

\begin{figure}
\begin{center}
\includegraphics[width=0.7\textwidth]{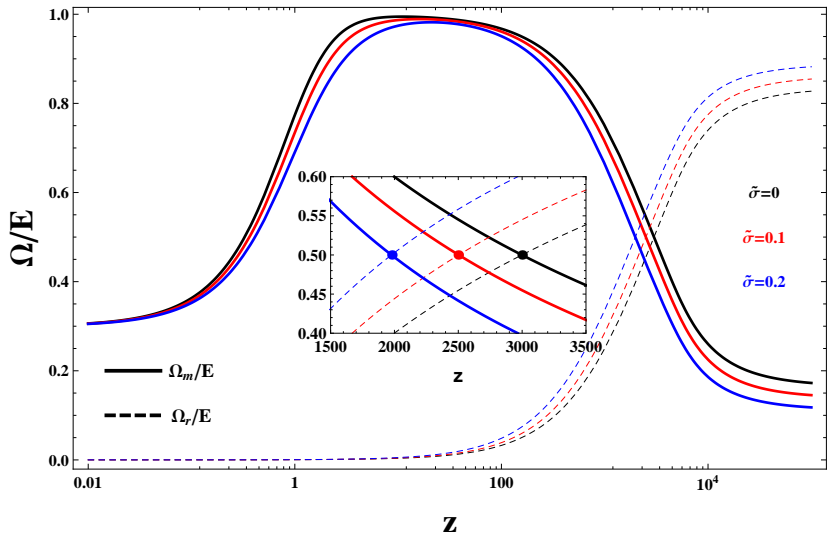}	
\caption{Evolution of the matter and radiation fractional densities. The initial data are $\Omega_{\phi0}=0.7$, $\Omega_{m0}=0.2999$, $\Omega_{r0}=1\cdot 10^{-4}$. The universe becomes radiation dominated at $z\sim 3000$ for $\sigma=0$ ($\Lambda$CDM), at $z\sim 2500$ for $\sigma=0.1$ and at $z\sim 2000$ for $\sigma=0.2$.}
\label{figrad}
\end{center}
\end{figure}

The $\phi$CDM model does not take into account the effect of the radiation component of the universe. To introduce this effect we make the fundamental assumption that radiation, in contrast to matter, does {\it not} undergo diffusion in the scalar field. From the point of view of the model this means that the energy-momentum tensor $T^\mathrm{rad}_{\mu\nu}$ of the radiation component is conserved: $\nabla^\mu T^{\mathrm{rad}}_{\mu\nu}=0$. In a spatially flat RW metric, the latter equation entails
\begin{equation}\label{omegarad}
\Omega_r(z)=\Omega_{r0}(1+z)^4,
\end{equation}
where $\Omega_r$ is the normalized energy density of radiation. The equations for the normalized densities of matter and dark energy remain the same, i.e., Eqs.~\eqref{dimensionlesseq1new}-\eqref{dimensionlesseq2new}, however the definition of $E(z)$ changes from~\eqref{dimensionlesseq3new} to
\begin{equation}\label{Erad}
E(z)=\sqrt{\Omega_m(z)+\Omega_\phi(z)+\Omega_r(z)},
\end{equation}
where $\Omega_r(z)$ is given by~\eqref{omegarad}. The behavior of solutions to the system~\eqref{dimensionlesseq1new},~\eqref{dimensionlesseq2new},~\eqref{omegarad},~\eqref{Erad} is depicted in Fig.~\ref{figrad}. On can see in particular that also the diffusion model under study, as the $\Lambda$CDM model, predicts that the universe was radiation dominated at early time. However according to the $\phi$CDM model the switch from a radiation-dominated to a matter-dominated universe is a more recent event than predicted by the $\Lambda$CDM model. 

\subsection{Cosmological perturbations of spatially flat solutions}\label{cosmopert}
Our next purpose is to study the evolution of linear scalar perturbations on the spatially flat Robertson-Walker metric~\eqref{spatiallyflatRW}. For an introduction to cosmological perturbations theory, see~\cite{malik,mukha}. Using the conformal Newtonian gauge in the absence of anisotropic stresses, we write the metric in the form
\begin{equation}\label{metric}
g=a(\eta)^2[-(1+2\Psi)d\eta^2+(1-2\Psi)\delta_{ij}dx^i dx^j]=\overline{g}+\delta g,
\end{equation} 
where $\Psi=\Psi(\eta,x^1,x^2,x^3)$ is the Newtonian potential, 
\[
\overline{g}=a(\eta)^2(-d\eta^2+\delta_{ij}dx^idx^j)
\]
is the spatially flat RW metric in the conformal time $\eta$ and 
\[
\delta g=-2a(\eta)^2\Psi (d\eta^2+\delta_{ij}dx^idx^j)
\]
is the (scalar) metric 
perturbation. Our convention is that, for any function $Q$, the evaluation of $Q$ on the background is denoted by  $\overline{Q}$ and the linear perturbation by $\delta Q$, i.e., 
\[
Q=\overline{Q}+\delta Q+O(\delta Q^2).
\]
We emphasize that barred quantities depend only on the conformal time $\eta$.

The components of Einstein's tensor for the metric~\eqref{metric} read (neglecting second order terms)
\[
G^0_{\ 0}=\overline{G^0_{\ 0}}+\delta G^0_{\ 0},\quad G^0_{\ i}=\overline{G^0_{\ i}}+\delta G^0_{\ i},\quad G^i_{\ j}=\overline{G^i_{\ j}}+\delta G^i_{\ j},
\] 
where, denoting $\mathcal{H}=(\log a(\eta))'$, $(\cdot)'=d(\cdot)/d\eta$,
\begin{equation}\label{G00}
\overline{G^0_{\ 0}}=-3\frac{\mathcal{H}^2}{a^2},\quad \delta G^0_{\ 0}=2a^{-2}[3\mathcal{H}(\mathcal{H}\Psi+\Psi')-\nabla^2\Psi],
\end{equation}
\begin{equation}\label{G01}
\overline{G^0_{\ i}}=0,\quad \delta G^0_{\ i}=2a^{-2}\partial_{x^i}(\mathcal{H}\Psi+\Psi').
\end{equation}
\begin{equation}\label{G11}
\overline{G^i_{\ j}}=-a^{-2}(\mathcal{H}^2+2\mathcal{H}')\delta^i_{\ j},\quad \delta G^i_{\ j}=2a^{-2}[\Psi''+3\mathcal{H}\Psi'+(2\mathcal{H}'+\mathcal{H}^2)\Psi]\delta^i_{\ j}.
\end{equation}
In the previous equations, $\partial_{x^i}$ denotes the partial derivative with respect to the variable $x^i$ and $\nabla^2$ is the flat Laplacian in the spatial variables, i.e., $\nabla^2=\sum_{i=1}^2\partial_{x^i}^2$. 

Next the perturbations of the scalar field and of the fluid variables will be considered. We set
\begin{equation}\label{pertur}
\phi=\overline{\phi}+\delta\phi,\quad \rho=\overline{\rho}+\delta\rho,\quad p=\overline{p}+\delta p,\quad n=\overline{n}+\delta n.
\end{equation}
As to the four-velocity, we have $u_\mu=\overline{u_\mu}+\delta u_\mu$, where
\[
\overline{u_\mu}=-a(t)\delta^0_{\ \mu}.
\]
Since at first order
\[
g^{\mu\nu}u_\mu u_\nu=-1+2(\Psi+a^{-1}\delta u_0),
\]
the requirement that $g^{\mu\nu}u_\mu u_\nu=-1$ hold at the first order entails
\[
\delta u_0=-a\Psi.
\]
Moreover it will be shown below (see the remark following Eq.~\eqref{phieq1}) that diffusion in a scalar field restricts the velocity perturbations to be of the form $\delta u_i=\partial_{x^i}
V$, for some scalar function $V$. It is convenient to set $V=a(\eta)\theta(\eta,x^1,x^2,x^3)$. In conclusion
\begin{equation}\label{ucov}
u_0=-a\,(1+\Psi),\quad u_i=a\,\partial_{x^i}\theta
\end{equation}
and therefore
\begin{equation}\label{u}
u^0=a^{-1}(1-\Psi),\quad u^i=a^{-1}\partial_{x^i}\theta.
\end{equation}
Substituting~\eqref{pertur},~\eqref{ucov} and~\eqref{u} into the definition of $T_{\mu\nu}$, see~\eqref{TandJfluid}, we obtain
\begin{subequations}
\begin{align}
&T^0_{\ 0}=\overline{T^0_{\ 0}}+\delta T^0_{\ 0},\quad \overline{T^0_{\ 0}}=-\overline{\rho},\ \delta T^{0}_{\ 0}=-\delta\rho,\\
&T^0_{\ i}=\overline{T^0_{\ i}}+\delta T^0_{\ i}
,\quad \overline{T^0_{\ i}}=0,\ \delta T^{0}_{\ i}=(\overline{\rho}+\overline{p})\partial_i\theta,\\
&T^i_{\ j}=\overline{T^i_{\ j}}+\delta T^i_{\ j},\quad \overline{T^i_{\ j}}=\overline{p}\,\delta^i_{\ j},\ \delta T^{i}_{\ j}=\delta p\,\delta^i_{\ j}.
\end{align} 
\end{subequations}
It follows that the Einstein equations~\eqref{EinsteinEq} at zero order, i.e., $\overline{G^\mu_{\ \nu}}+\overline{\phi}\,\delta^\mu_{\ \nu}=\overline{T^\mu_{\ \nu}}$, read 
\begin{equation}\label{einsteinback}
3\frac{\mathcal{H}^2}{a^2}=\overline{\rho}+\overline{\phi},\quad \mathcal{H}'=-\frac{1}{2}[a^2(\overline{p}-\overline{\phi})+\mathcal{H}^2],
\end{equation}
while at first order, i.e., $\delta G^\mu_{\ \nu}=\delta T^\mu_{\ \nu}$, they give
\begin{align}
&\nabla^2\Psi-3\mathcal{H}(\mathcal{H}\Psi+\Psi')=\frac{1}{2}a^2(\delta\rho+\delta\phi),\label{einstein1ham}\\
&\partial_{x^i}(\mathcal{H}\Psi+\Psi')=\frac{1}{2}a^2(\overline{\rho}+\overline{p})\partial_{x^i}\theta\Rightarrow \mathcal{H}\Psi+\Psi'=\frac{1}{2}a^2(\overline{\rho}+\overline{p})\theta,\label{einstein1vec}\\
&\Psi''+3\mathcal{H}\Psi'+(2\mathcal{H}'+\mathcal{H}^2)\Psi=\frac{1}{2}a^2(\delta p-\delta\phi),\label{einstein1ten}
\end{align}
where in the second version of~\eqref{einstein1vec} we set an arbitrary function of $\eta$ equal to zero.

Now Eq.~\eqref{Jeqfluid} at zero order gives
\begin{equation}\label{neq0}
\overline{n}'+3\mathcal{H}\overline{n}=0\Rightarrow \overline{n}(\eta)=\overline{n}_0\,a(\eta)^{-3},
\end{equation}
where $\overline{n}_0=\overline{n}(0)$ and $a(0)=1$.
At first order we obtain
\begin{equation}\label{neq1}
\delta n'+3\mathcal{H}\delta n+\overline{n}\nabla^2\theta-3\overline{n}\Psi'=0.
\end{equation}
The equation~\eqref{phieqfluid} for the scalar field gives, at zero order,
\begin{equation}\label{phieq0}
\overline{\phi}'=-\sigma a \overline{n},
\end{equation}
and at first order
\begin{equation}\label{phieq1}
\delta \phi'=-\sigma a(\delta n+\overline{n}\Psi)\quad \partial_{x^i}\delta \phi=\sigma a\overline{n}\partial_{x^i}\theta\Rightarrow\delta\phi=\sigma a\overline{n}\theta.
\end{equation}
{\bf Remark:} If we had chosen a general perturbation $\delta u_i$ for the spatial velocity, the scalar field equation at first order would have given $\partial_{x^i}\phi=\sigma\overline{n}\delta u_i$, by which it follows again that $\delta u_i=a\partial_{x^i}\theta$. Hence our original choice for $\delta u_i$ entails no loss of generality. To put it differently, at first order the fluid velocity is irrotational.

Note that $\delta\phi$ is completely determined by the perturbation $\theta$. Substituting $\delta\phi=\sigma a\overline{n}\theta$ in the first of~\eqref{phieq1}, we obtain $2\mathcal{H}\overline{n}\theta-\overline{n}\theta'=(\delta n+\overline{n}\Psi)$, which can be used to express $\delta n$ in terms of the perturbations $\theta$ and $\Psi$.

Using that $\overline{n} a^3=\overline{n}_0$,  the equations for the independent perturbations $\Psi,\theta,\delta\rho,\delta p$ become
\begin{subequations}\label{gensys}
\begin{align}
&\nabla^2\Psi-3\mathcal{H}(\mathcal{H}\Psi+\Psi')=\frac{a^2}{2}\delta\rho+\frac{1}{2}\sigma \overline{n}_0\theta,\label{Psieq1}\\
& \mathcal{H}\Psi+\Psi'=\frac{1}{2}a^2(\overline{\rho}+\overline{p})\theta,\label{Psieq2}\\
&\Psi''+3\mathcal{H}\Psi'+(2\mathcal{H}'+\mathcal{H}^2)\Psi=\frac{a^2}{2}\delta p-\frac{1}{2}\sigma \overline{n}_0\theta.\label{phi2eq}
\end{align}
\end{subequations}
To close the system we need an equation of state, which we take to be that of a dust fluid.
Setting $\overline{p}=\delta p=0$  and using the second equation in~\eqref{einsteinback} in the system~\eqref{gensys} we obtain
\begin{subequations}\label{gensysp=0}
\begin{align}
&\nabla^2\Psi-3\mathcal{H}(\mathcal{H}\Psi+\Psi')=\frac{a^2}{2}\delta\rho+\frac{1}{2}\sigma \overline{n}_0\theta,\\
& \mathcal{H}\Psi+\Psi'=\frac{1}{2}a^2\overline{\rho}\theta,\\
&\Psi''+3\mathcal{H}\Psi'+a^2\overline{\phi}\,\Psi=-\frac{1}{2}\sigma\overline{n}_0 \theta,
\end{align}
\end{subequations}
where $a,\overline{\rho},\overline{\phi},\mathcal{H}$ solve
\begin{equation}\label{backp=0}
a'=a\mathcal{H},\quad \mathcal{H}'=\frac{a^2}{2}\overline{\phi}-\frac{\mathcal{H}^2}{2},\quad \overline{\phi}'=-\frac{\sigma \overline{n}_0}{a^2},\quad \overline{\rho}'+3\mathcal{H}\overline{\rho}=\frac{\sigma \overline{n}_0}{a^2},\quad 3\frac{\mathcal{H}^2}{a^2}=\overline{\rho}+\overline{\phi}
\end{equation}
with initial data $a(0)=1$, $\mathcal{H}_0>0$, $\overline{\rho}_0>0$, $\overline{\phi}_0=3\mathcal{H}_0^2-\overline{\rho}_0>0$.
Combining the second and the third equation of the system~\eqref{gensysp=0} we obtain an equation for $\Psi$ alone:
\begin{equation}\label{eqpsifin}
\Psi''+\left(3\mathcal{H}+\frac{\sigma\overline{n}_0}{a^2\overline{\rho}}\right)\Psi'+\left(a^2\overline{\phi}+\mathcal{H}\frac{\sigma \overline{n}_0}{a^2\overline{\rho}}\right)\Psi=0.
\end{equation}
 
\subsection{Solutions of the linearized equations}\label{sollin}
Once we solve~\eqref{eqpsifin} for given initial data $\Psi(0,x)$, $\Psi'(0,x)$, we replace the solution in the first two equations of the system~\eqref{gensysp=0} and determine the perturbations $\delta\rho$, $\theta$, and so in particular the density contrast defined by
\[
\Delta_m=\frac{\delta\rho}{\overline{\rho}}.
\] 
Of course this procedure can in general only be performed numerically. However, as a way of example, we show next how it works for perturbations around the exact solution~\eqref{solution}$_{K=0,\gamma=1}$ of the background equations, which in conformal time reads
\[
a(\eta)=e^{\mathcal{H}_0\eta},\quad\mathcal{H}(\eta)=\mathcal{H}_0,\quad \overline{\phi}(\eta)=\overline{\phi}_0a(\eta)^{-2},\quad \rho(\eta)=\overline{\rho}_0a(\eta)^{-2},
\]
where
\[
\mathcal{H}_0=\left(\frac{\sigma \overline{n}_0}{2}\right)^{1/3},\quad \overline{\phi}_0=\left(\frac{\sigma \overline{n}_0}{2}\right)^{2/3},\quad\overline{\rho}_0=(\sqrt{2}\sigma \overline{n}_0)^{2/3}.
\]
For this solution, equation~\eqref{eqpsifin} becomes
\begin{equation}\label{psieqspec}
\Psi''+4\mathcal{H}_0\Psi'+2\mathcal{H}_0^2\Psi=0.
\end{equation}
The solution of the previous equation is
\begin{equation}\label{solspec}
\Psi(\eta,x)=C_-(x)e^{-(2+\sqrt{2})\mathcal{H}_0\eta}+C_+(x)e^{-(2-\sqrt{2})\mathcal{H}_0\eta},
\end{equation}
where $C_\pm(x)$ are arbitrary functions fixed by the initial data. Precisely, letting
\[
\Psi_\mathrm{in}(x)=\Psi(0,x),\quad \dot{\Psi}_\mathrm{in}(x)=\Psi'(0,x),
\]
we find
\begin{equation}\label{indata}
C_+(x)=\Psi_\mathrm{in}(x)\left(\frac{1}{2}-\frac{1}{\sqrt{2}}\right)-\frac{\dot{\Psi}_\mathrm{in}(x)}{2\sqrt{2}\mathcal{H}_0},\quad C_-(x)=\Psi_\mathrm{in}(x)\left(\frac{1}{2}+\frac{1}{\sqrt{2}}\right)+\frac{\dot{\Psi}_\mathrm{in}(x)}{2\sqrt{2}\mathcal{H}_0}.
\end{equation}

\begin{figure}
\begin{center}
\includegraphics[width=0.485\textwidth]{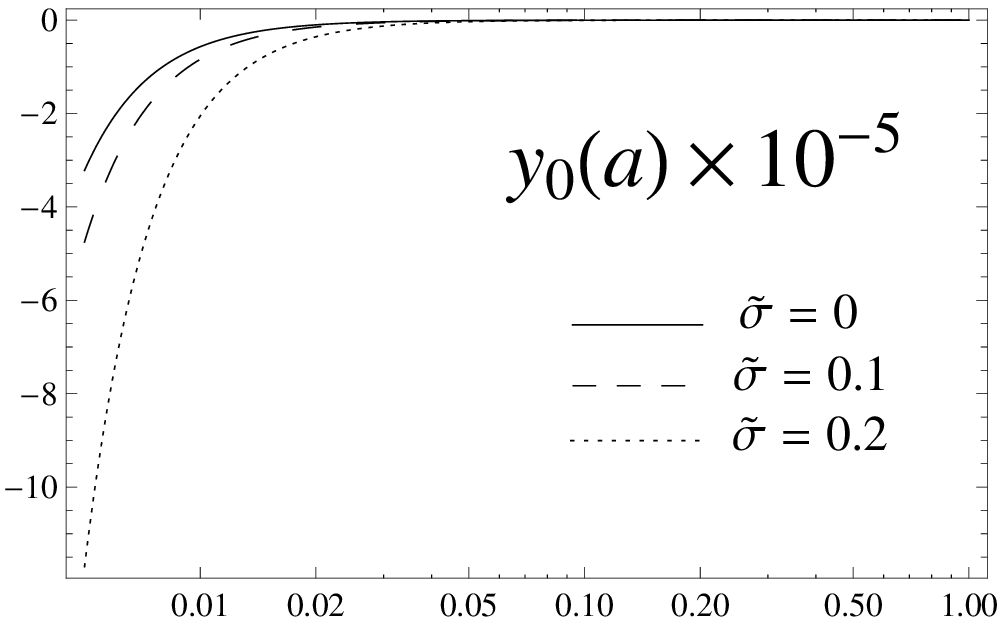}\quad
\includegraphics[width=0.485\textwidth]{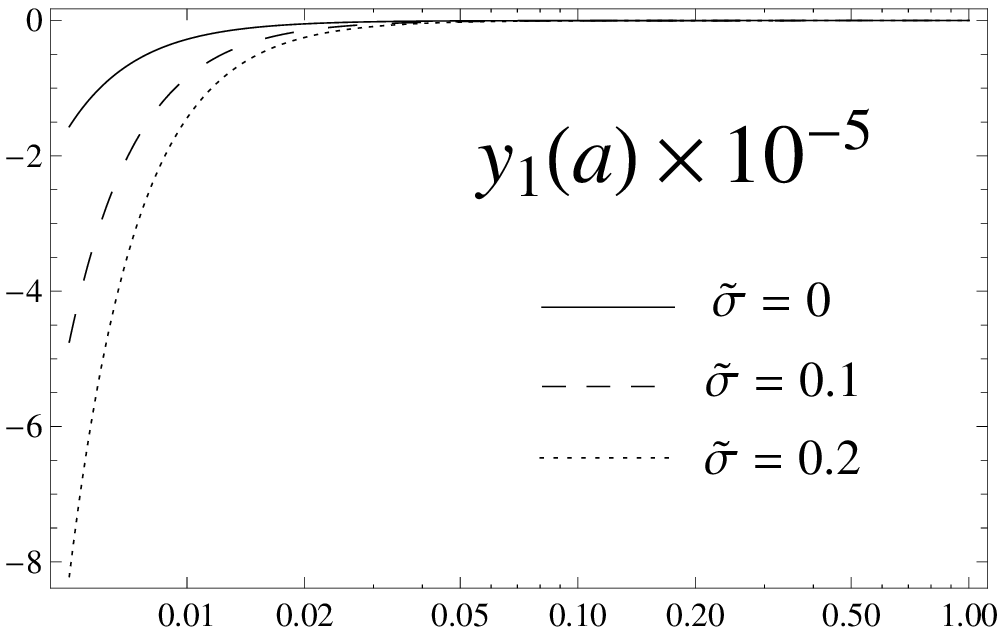}	
\caption{Evolution of the linear scalar gravitational fluctuations as functions of the scalar factor (assuming $\Omega_{m0}=0.3$). }
\label{y0y1}
\end{center}
\end{figure}

In general, if we denote by $y_0(\eta), y_1(\eta)$ the solutions of the ODE
\begin{equation}\label{eqpsifinODE}
y''+\left(3\mathcal{H}+\frac{\sigma\overline{n}_0}{a^2\overline{\rho}}\right)y'+\left(a^2\overline{\phi}+\mathcal{H}\frac{\sigma \overline{n}_0}{a^2\overline{\rho}}\right)y=0
\end{equation}
with initial data 
\begin{equation}
y_0(0)=0,\ y'_0(0)=1;\quad y_1(0)=1,\ y_1'(0)=0,
\end{equation}
the general solution of~\eqref{eqpsifin} with initial data $\Psi_\mathrm{in}$, $\dot{\Psi}_\mathrm{in}$ is given by
\begin{equation}
\Psi(\eta,x)=\Psi_\mathrm{in}(x) y_1(\eta)+\dot{\Psi}_\mathrm{in}(x)y_0(\eta).
\end{equation}
In Fig.~\ref{y0y1} we show the numerical evolution of the functions $y_{0}$ and $y_1$ in terms of the scalar factor. The dashed lines correspond to the $\Lambda$CDM model, while continued lines refer to $\phi$CDM with $\sigma=0.1$ and $\sigma=0.2$. The difference in the evolution of the potential fluctuations results in important observable deviations of the two models, such as the anisotropy in the CMB temperature due to the integrated Sachs-Wolfe effect, which will be discussed in Section~\ref{ISWeffect}.

\section{Comparison with observations}\label{CompObs}
The purpose of this section is to test the viability of the $\phi$CDM model based on the available observations. A particular interest here is to constrain the maximal value of the diffusion coefficient $\tilde{\sigma}$ allowed by current data.

\subsection{Constraining the background expansion with data}\label{CompObsBack}

In order to compare the $\phi$CDM model with the data we have to obtain the expansion H(z) by solving numerically equations~\eqref{dimensionlesseq1}-\eqref{dimensionlesseq3} with appropriate initial conditions $\Omega_{m}(z=0)=\Omega_{m0}$ and $\Omega_{\phi}(z=0)=\Omega_{\phi_0}=1-\Omega_{m0}$. We constrain the model parameters with the following observational data sets. 

\noindent {\bf Supernovae}: First, we use Supernovae data from the recent UNION2.1 compilation \cite{suzuki}. This test is based on the observed distance modulus $\mu^{obs}(z)$ of each SN Ia at a certain redshift $z$,
\begin{equation}
\mu(z)=25+5 log_{10}\frac{d_L(z)}{Mpc}, 
\end{equation}
where the luminosity distance, in a spatially flat RW metric, is given by the formula
\begin{equation}
d_L(z)=c(1+z)\int^{z}_0\frac{dz^{\prime}}{H(z^{\prime})}. 
\end{equation}
The UNION 2.1 compilation provides $N_{SN}=580$ distance modulus $\mu_{i}(z_i)$ for different redshifts $z_i$.

\noindent{\bf Differential Age}:
A second observational source comes from the evaluation of the age of old galaxies that have evolved passively giving rise to the differential age data of such objects \cite{Ji1, Ji2, Hz}. Since the expansion rate is defined as
\begin{equation}
H(z)=-\frac{1}{1+z}\frac{dz}{dt}.
\end{equation}
Since spectroscopic redshifts of galaxies are known with very high accuracy, one just needs a differential measurement of time $dt$ at a given redshift interval in order to obtain values for H(z). The data used in this work consist on $N_{H}=21$ data points listed in \cite{Farooq}, but previously compiled in \cite{moresco}.

\noindent {\bf Baryon Acoustic Oscillations}:
The baryon acoustic oscillation (BAO) scale is calculated by the $D_V$ parameter
\begin{equation}
D_V(z)=\left[(1+z)^2 D_A^2(z)\frac{c z}{H(z)}\right]^{1/3},
\end{equation}
where $D_A(z)$ is the angular-diameter distance. The values for $D_V$ have been reported in the literature by several galaxy surveys. In our analysis we use data ate $z=0.2$ and $z=0.35$ from the SDSS \cite{sdss}, data at $z=0.44,0.6$ and $0.73$ from the WiggleZ \cite{wiggle} and one data point at $z=0.106$ from the 6dFGRS \cite{6dfgrs} surveys. In the total we have $N_{BAO}=6$.

For our statistical analysis we construct the chi-square function
\begin{equation}
\chi^2=\sum^{N}_{i=1}\frac{\left(f^{th}(z_i)-f^{obs}(z_i)\right)^2}{\sigma^2_i},
\end{equation}
where $f=\left(\mu,H,D_V\right)$ for the SN, H(z) and BAO datasets, respectively. The number of data points $\left\{z_i,f^{obs}(z_i)\right\}$ in each set is, respectively, $N=N_{SN}, N_{H}$ and $N_{BAO}$ 
whereas $\sigma_i$ is the observational error associated to each observation $f^{obs}$ and $f^{th}$ is the theoretical value predicted by the $\phi$CDM model.  

\begin{figure}
\begin{center}
\includegraphics[width=0.295\textwidth]{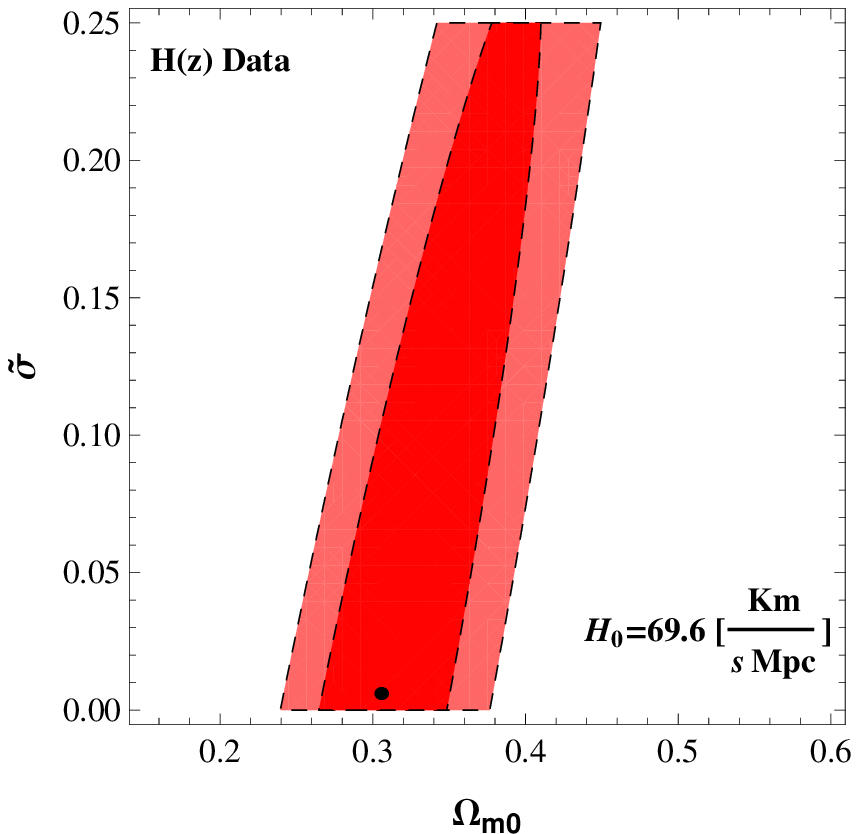}
\includegraphics[width=0.295\textwidth]{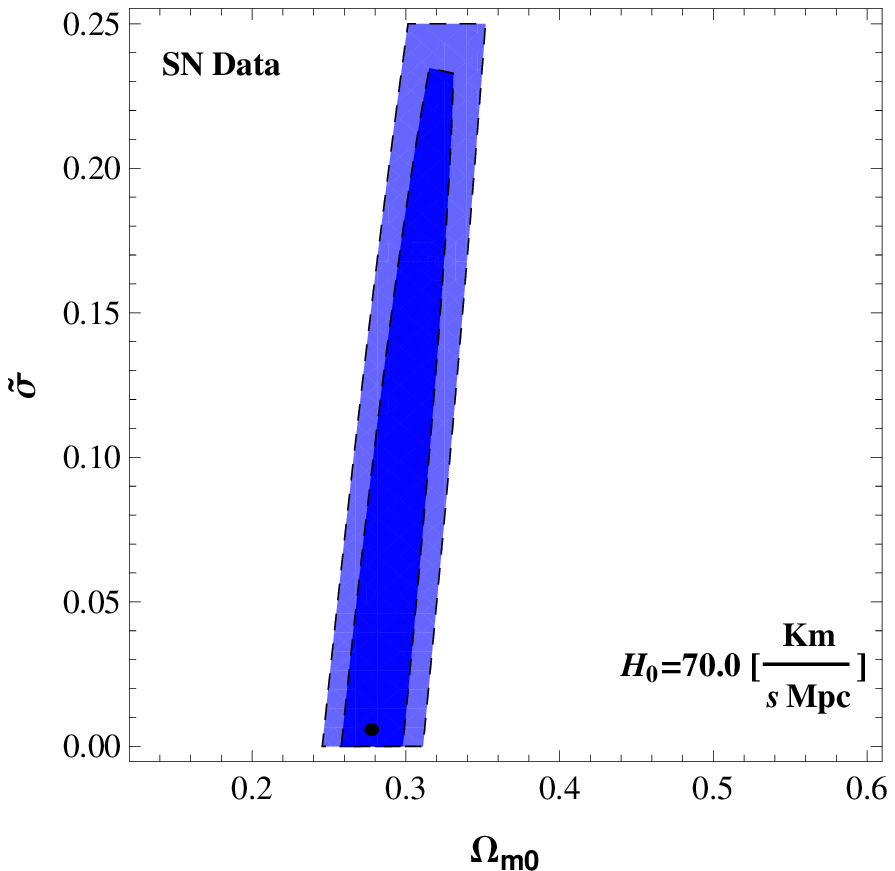}
\includegraphics[width=0.295\textwidth]{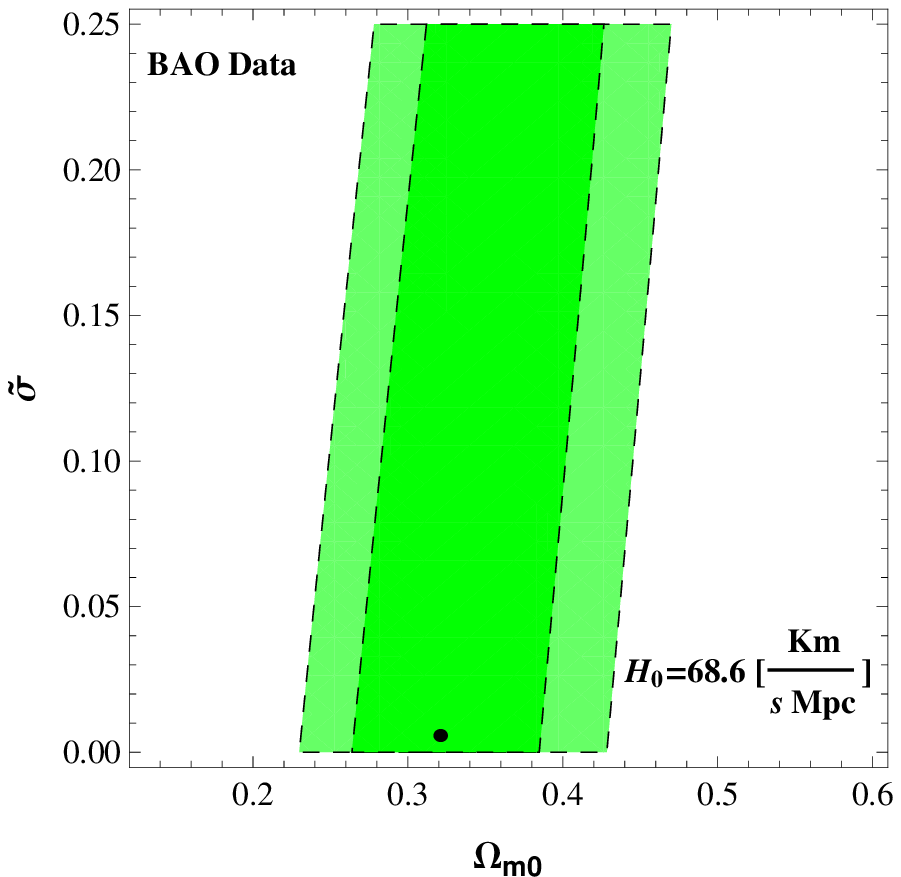}	
\caption{Confidence level contours at $1\sigma$ and $2\sigma$ for H(z) (Red), Supernovae (Blue), BAO (Green) and data sets. In each panel the $H_0$ value has been fixed following the best fit of the $\Lambda$CDM model for each sample.}
\label{fig2}
\end{center}
\end{figure}

Our diffusion $\phi$CDM model has 3 free parameters, namely $H_0, \Omega_{m0}$ and $\tilde{\sigma}$. We are mostly interested in the constraints on the diffusion parameter $\tilde{\sigma}$. We will fix delta priors on $H_0$ based on the results of the $\Lambda$CDM model for each data set individually. Having the $\Lambda$CDM model only $\left\{H_0, \Omega_{m0}\right\}$ as free parameters the H(z), UNION 2.1 SN and the BAO data place, individually, the best fit at $\left\{H_0, \Omega_{m0}\right\}=\left\{69.6,0.31\right\}$, $\left\{70.0,0.28\right\}$ and $\left\{68.6, 0.32\right\}$, respectively. The $H_0$ values obtained in this way (69.6,70.00,68.6) will be used in our analysis for the diffusion model as delta priors for the H(z), SN and BAO samples, respectively. This can be seen in Fig.~\ref{fig2}. With this strategy we avoid that $H_0$ becomes a possible source of degenerescence with the $\Lambda$CDM model. However, we note that our final conclusions are very weakly dependent on the prior choice for $H_0$. 

\begin{figure}
\begin{center}
\includegraphics[width=0.55\textwidth]{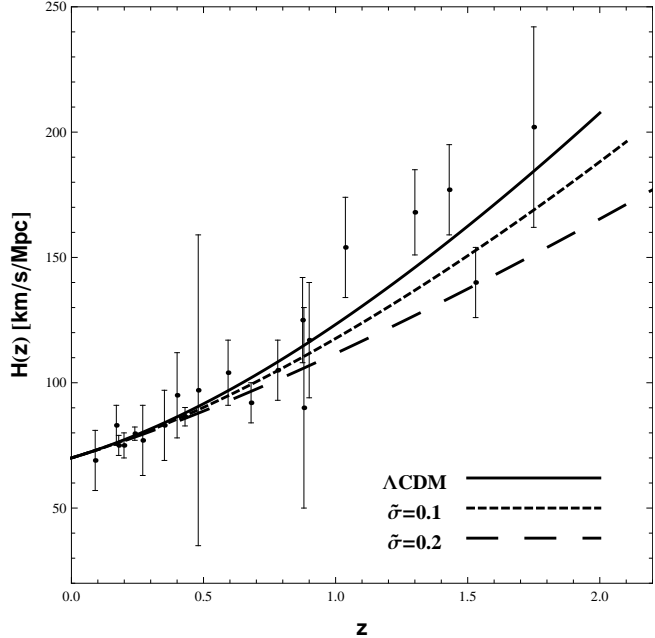}
\caption{Expansion H(z) versus the redshift. The data points shown correspond to the differential age of galaxies and are described in the text.}
\label{figH}
\end{center}
\end{figure}

In Fig.~\ref{fig2} we show the confidence level contours for our diffusion model on basis of 2 free parameters ($\tilde{\sigma}$ and $\Omega_{m0}$), i.e. $\Delta\chi^2=\chi^2-\chi^2_{min} < 2.30 (68 \%$ of confidence level) and $6.17 (95 \% )$. In this contours, the parameter $\tilde{\sigma}$ is limited to the value $\tilde{\sigma}=0.25$ since for larger values the dynamics shows bouncing solutions. As one can see in Fig.~\ref{figH} the $H(z)$ data is not well described for $\tilde{\sigma}>0.2$. The best-fit obtained for each data set is shown as a filled circle and corresponds to the $\Lambda$CDM model ($\tilde{\sigma}=0$). However, it is worth noting from Fig.~\ref{fig2} that SN, H(z) and BAO are not able to set an upper limit on $\tilde{\sigma}$, i.e. all values of $\tilde{\sigma}$ that produce viable solutions are consistent at $95\%$ of CL with background observations. 

In order to find stronger constraints on $\tilde{\sigma}$ we study in the next subsection the structure formation process.

\subsection{Structure Formation} \label{CompObsPert}

Quantum fluctuations produced during the inflationary epoch are responsible for the origin of the structures like galaxies and clusters of galaxies that we observe. These small fluctuations evolve under gravitational interaction until a final stage where they become highly nonlinear collapsed objects. However, the study of the linear regime of the cosmological perturbation theory, i.e. when the perturbed quantities have amplitudes much smaller than their background values, allows us to understand the main aspects of structure formation at very large scales. For example, the main patterns of observations like the cosmic microwave background (CMB) and the statistical distribution of the matter density field at large scales can be fully described by the linear regime.

\subsubsection{Cosmic microwave background anisotropies}\label{ISWeffect}

The temperature anisotropies on the CMB sky are connected to the linear fluctuations of matter via the Sachs-Wolfe effect \cite{ISW}. While most of such anisotropies were already present at the time of last scattering  ($z\sim 1100$), a relevant part of them has been produced thereafter due to the fact that photons traveled through time varying gravitational potential wells. The latter contribution, known as the integrated Sachs-Wolfe effect (ISW), can be computed as the integral of the derivative of the potential fluctuations along the photon trajectory $\hat{{\bf n}}$ from $\eta_{\rm lss}$ (conformal time
at the last scattering surface or decoupling time) to $\eta_0$ (conformal time today). For the ISW only, and for the case where there is no shear stress, we can write the CMB temperature anisotropy as

\begin{equation}\label{deltaTISW}
\left(\frac{\Delta T}{T}\right)_{\rm
ISW}=2\int^{\eta_{0}}_{\eta_{lss}}d\eta\frac{\partial\Psi}{\partial
\eta}\left[\left(\eta_{0}-\eta\right){\bf \hat{ n}},\eta\right].
\end{equation}

As we saw in Section~\ref{sollin}, even assuming that the potential fluctuations at redshift $z_{lss}$ are the same for the $\Lambda$CDM model and the $\phi$CDM model, diffusion alters considerably the time evolution of the gravitational perturbations, leading to a possible different contribution to the ISW effect. In order to show this, in Fig.~6 we plot the temporal evolution of the gravitational potential comparing the $\Lambda$CDM with the diffusion model for some relevant modes. The amplitudes for each mode can be easily calculated with the CAMB code \cite{lewis}. We are interested on very large scales $k_{ISW} \geq 0.0005 Mpc^{-1}$ where the CMB spectrum is most sensitive to the ISW effect. As seen in Fig.~6 the value $\tilde{\sigma}=0.1$ is capable to modify the evolution of the gravitational potential at late times. This behavior has important consequences for the ISW.

\begin{figure}	
\begin{center}
\includegraphics[width=0.55\textwidth]{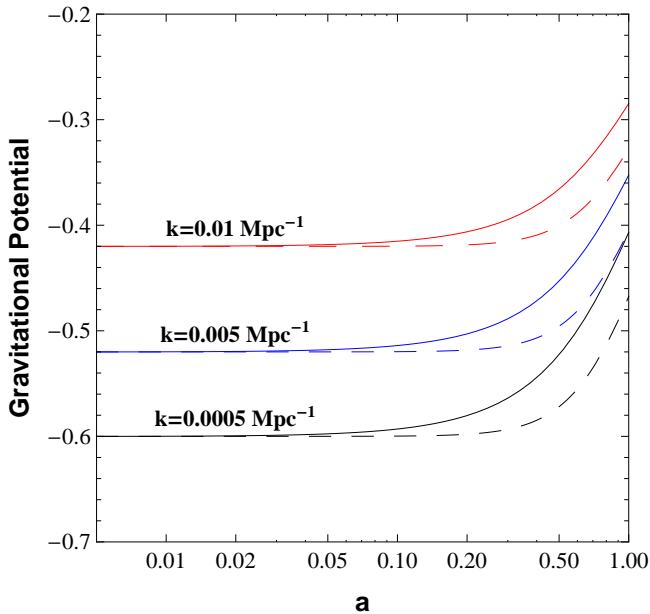}
\caption{Gravitational potential for three different scales as a function of the scale factor for the $\Lambda$CDM (dashed) and the diffusion model with $\tilde{\sigma}=0.1$ (solid).}
\end{center}
\label{fig4}
\end{figure}

Focusing on the computation of the ISW effect, we employ the same strategy as proposed in \cite{dent}. With the perturbed equations deduced before we can calculate the ISW signal Eq. (\ref{deltaTISW}). A comparison between the $\Lambda$CDM model and the diffusion one is performed by calculating relative amplifications ($Q$) of the ISW effect as
\begin{equation}\label{Q}
Q\equiv\frac{\left(\frac{\Delta T}{T}\right)_{\rm ISW}^{\rm Diff}}{\left(\frac{\Delta T}{T}\right)_{\rm ISW}^{\rm \Lambda CDM}}-1.
\end{equation}
If $Q>0 \,(<0)$ the diffusion model produces more (less) temperature variation to the CMB photons via the ISW effect than the fiducial $\Lambda$CDM model.  

Assuming a fiducial $\Lambda$CDM model with $\Omega_{m0}=0.30$ and $h=0.7$ we calculate the quantity $\left(\frac{\Delta T}{T}\right)_{\rm ISW}^{\rm \Lambda CDM}$ using the $\Lambda$CDM equations, i.e. the case $\tilde{\sigma}=0$. Then, we calculate $\left(\frac{\Delta T}{T}\right)_{\rm ISW}^{\rm Diff}$ for many different values of the parameters $\tilde{\sigma}$ and $\Omega_{m0}$. Fig.~7 shows the values of the relative amplification $Q$ in the plane $\tilde{\sigma}$ x $\Omega_{m0}$.

One can see that, for instance, if $\Omega_{m0}=0.3$ and $\tilde{\sigma}=0.1$ the diffusion model shows an enhancement of order $80\%$. Although the $\Lambda$CDM provides a good fit to the CMB data, note that the ISW occurs at the largest scales where the cosmic variance introduces large errors bars on the CMB measurements. To obtain pure ISW data is a complex task since it demmands a cross correlation between the CMB maps and the large scale structure catalogues. Our goal here was to keep the ISW signal of the diffusion model under control avoiding the parameters that produces a huge amplification of this signal. The question seems to be what are the unacceptable values for $Q$. Some recent analysis of the stacked ISW signal of superstructures present in the SDSS DR6 MegaZ catalogue \cite{granett} report a observed value of $\Delta T=9.6 \pm 2.2 \mu K$ which is almost five times larger than the $\Lambda$CDM theoretical prediction $\Delta T=2.27 \pm .014 \mu K$ \cite{sesh}. This corresponds to $Q$ values of order $\sim 400\%$, which allows for larger values of the parameter $\tilde{\sigma}$ than considered so far, and indicates that models with amplified ISW signals could be favored over the $\Lambda$CDM model, but this is currently an open issue.

\begin{figure}
\begin{center}
\includegraphics[width=0.55\textwidth]{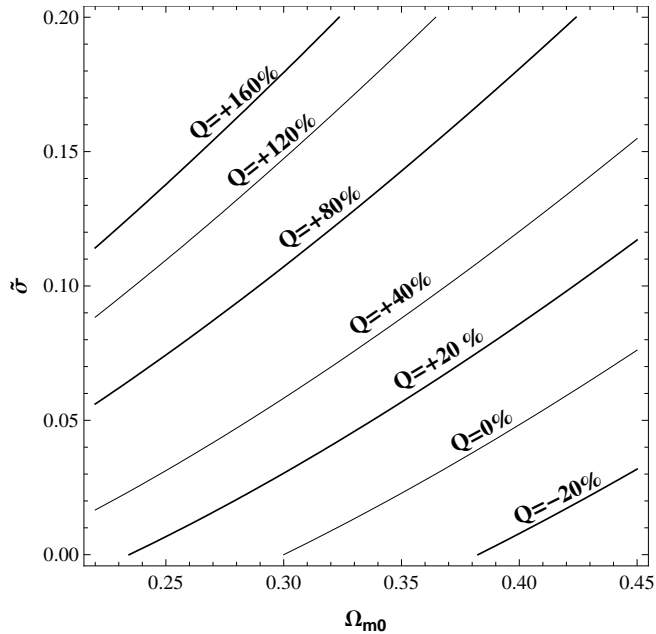}
\label{fig5}	
\caption{Contours for the parameter $Q$ in the $\tilde{\sigma}$ x $\Omega_{m0}$ plane.}  
\end{center}
\end{figure}

\subsubsection{The matter power spectrum}

Large scale galaxy surveys like the 2dFGRS, SDSS and the WiggleZ projects observe the clustering properties of galaxies providing the correlation function in the space of the position $\xi(r)= \left\langle \Delta(r) \Delta(x+r) \right\rangle$. Its Fourier transform is the matter power spectrum $P(k,z)$ in the wavenumber space. It can be computed at any time $z$, but the observations provides today's power spectrum which is defined as
\begin{equation}
P(k)=\left|\Delta(k, z=0)\right|^2.
\end{equation}

Our procedure to obtain the power spectrum is the following. We first assume a primordial post-inflationary power spectrum $P_i\sim k^{n_s}$ with spectral index $n_s=0.96$ \cite{planck16}. Then we obtain today's power spectrum by evolving the primordial one until a redshift $z=0$ using the BBKS transfer function \cite{bbks}. This provides the power spectrum for the standard $\Lambda$CDM model which can be seen as the solid line in Fig.~8. This spectrum provides a good fit to the data. Here, we show the data points obtained by the two degree field galaxy redshift survey (2dFGRS) \cite{2dFGRS} for illustrative reasons. Since we are not planning to performe a statistical analysis it is not necessary to use more recent data.

Returning to the procedure for obtaining the spectrum, the next step is to integrate back in time the perturbed $\Lambda$CDM equations in order to find the initial conditions for $\Delta$ at some point during the matter dominated epoch, let's say $z_i=500$. At this redshift we expect that our diffusion model behaves exactly as the standard model meaning that they have the same initial conditions in the past. Now, we use the power spectrum calculated at $z_i=500$ as initial condition for the diffusion model. Finally, we integrate the diffusion equation from $z_i$ until $z=0$ assuming different values of $\tilde{\sigma}$. We also normalize the spectrum in order to match the data at small scales, which corresponds to a $\sigma_{8}$ normalization. This procedure provides the matter power spectrum for the diffusion $\phi$CDM model.

In Fig.~8 we compare the matter power spectrum for the $\Lambda$CDM and the diffusion models for some values of the parameter $\tilde{\sigma}$. 
The matter power spectrum provides a statistical measurement of the matter distribution. Its peak occurs at a scale $k_{eq}$ that depends on the redshift $z_{eq}$ of matter-radiation equality. Scales smaller than $k_{eq}$, i.e. $k>k_{eq}$, are subjected to sub-horizon damping before $z_{eq}$. Since the diffusion mechanism does not modify the early stages of evolution of the universe, i.e., a radiation dominated phase in the past, we expect the same small scale behavior and for this reason we normalize the power spectrum at $k=0.185~h~Mpc^{-1}$ (the smaller scale accessible with the linear theory). 

We see that the matter power spectrum is a very sensitive probe for the diffusion model and small values of $\tilde{\sigma}$ drastically modify its shape. In general, it shows a pronounciated power suppression at large scales (small values of $k$). Note that the peak of the spectrum is related to the moment of matter-radiation equality. As shown in Fig.~\ref{figrad} the transition redshift is also very sensitive to the $\tilde{\sigma}$ values and therefore the peak is shifted. This analysis seems to place much stronger constraints on the diffusion parameter $\tilde{\sigma}$ than the ones presented in the previous section. It is clear that values of the order $\tilde{\sigma}=0.1$ of cosmic diffusion are challenged by this analysis, whilst values $\tilde{\sigma}<0.01$ seem to be in agreement with large scale structure data. We observe that with such normalization, the power at large scales is strongly suppressed for values of order $\tilde{\sigma}\sim 0.1$. The shape of the spectrum is preserved only for values or order $\tilde{\sigma} \lesssim 0.01$ which represents a constrain one order of magnitude stronger than the previous analysis using background data only.

\begin{figure}
\begin{center}
\includegraphics[width=0.55\textwidth]{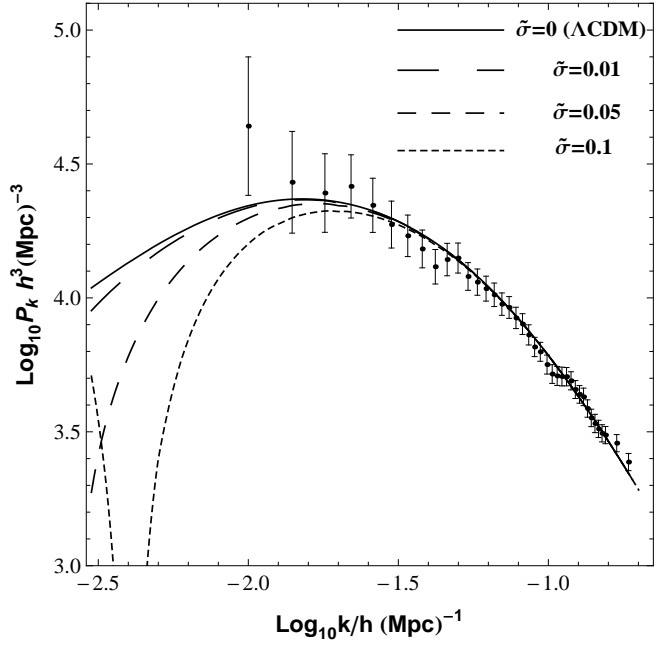}
\label{fig6}	
\caption{Matter power spectrum. The $\Lambda$CDM ($\tilde{\sigma}=0$) model is shown in the solid line. The data points correspond to the 2dFGRS data. The case of cosmic diffusion is plotted in the dashed lines for different values of the parameter $\tilde{\sigma}$.}
\end{center}
\end{figure}

\section{Conclusions}

We have constructed a viable general relativistic cosmological model where the microscopic velocity diffusion of the matter particles  is taken into account. The model is characterized by a single parameter $\sigma>0$ measuring the energy gained by the particles due to the action of the diffusion forces. The local conservation of energy (expressed through the Bianchi identities) demands the existence of an additional field in spacetime that transfer energy to the matter particles. We identified such field with the dark energy component of the universe and modeled it by a cosmological scalar field $\phi$ added to the Einstein equations. This leads to the $\phi$CDM model. The $\Lambda$CDM model is fully recovered in the absence of diffusion ($\sigma=0$). If diffusion takes place ($\sigma >0$), the density of the matter fluid is coupled to the density associated to the $\phi$ field. We have not included the radiation component for most of our analysis, but we have shown that if a relativistic fluid is included into the background expansion, then $\phi$CDM is radiation dominated at early times, becomes subsequently matter dominated and finally, only recently, enters in a phase of accelerated expansion dominated by the $\phi$ field. Thus the paradigm of the standard cosmological models is preserved.

We argue that the coupling caused by the diffusion process can be seen as a theoretical motivation for the interaction between dark matter and dark energy. Interacting cosmological models have been widely studied as an alternative to the standard $\Lambda$CDM scenario as a possible solution for the coincidence problem, but up to now they have been only phenomenologically inspired. As far as we know we are providing, through the diffusion mechanism, the first physically motivated justification for the interaction between dark matter and dark energy.

We have also used observational data in order to place constraints on the magnitude of the diffusion forces characterized by a normalized (dimensionless) parameter $\tilde{\sigma}$. We have found that bouncing solutions for the background expansion occur for values $\tilde{\sigma} \gtrsim 0.25$. Using the background data in order to place constraints on the viable cosmological scenarios with $\tilde{\sigma} \lesssim 0.25$ we have shown that Supernovae, H(z) and BAO data are not so restrictive. The background data is well described at least within a $95$\% of confidence level.

The most strong constraints appear from the analysis of the large scale structures formation. We have obtained solutions for the gravitational potential at large scales. The temporal evolution of the potential is very sensitive to the diffusion parameter. This led us to investigate how the integrated Sachs-Wolfe signal is affected. Indeed, values $\tilde{\sigma} \sim 0.1$ can produce an ISW effect which is of order $80 \%$ larger than the standard $\Lambda$CDM signal. However, this excess of power is not necessarily undesirable. Indeed, having in mind recent results from the cross-correlation between CMB maps and galaxy surveys, the observed ISW signal could be $~400\%$ larger than predicted by the standard model. It seems that, while one has to keep control on the magnitude of the ISW signal, reliable constraints on such quantity are still not available to the community. We leave for a future work the full analysis of the present model within the CMB data.

Our final analysis using the matter power spectrum data was the most successful in constraining the parameter $\tilde{\sigma}$. Our $\phi$CDM model is not ruled out by the LSS analysis, but the latter provides the stronger constraints on the magnitude of the diffusion of matter particles. We concluded that diffusion is still a viable process in the universe if the total matter undergoes diffusion with magnitude $\tilde{\sigma} \lesssim 0.01$. Note that our approach does not differentiate between baryonic and dark matter. The inclusion of a baryonic component in our model would produce a more realistic model. In this case, the forthcoming data from galaxy redshift surveys, in particular the surveys designed to measure the baryonic acoustic oscillations scale, could be much more restrictive in constraining the existence of matter diffusion effects in the universe.

\acknowledgments

HV is supported by CNPq (Brazil). We thank the organizers of the 49th Winter School of Theoretical Physics Cosmology and non-equilibrium statistical mechanics (Ladek-Zdr\'oj, Poland).

\end{document}